\documentclass{nature}

\usepackage{amsmath}
\usepackage{amssymb}
\usepackage{graphicx}
\usepackage{upgreek}
\usepackage{float}
\usepackage{bigints}

\def\arcmin{\hbox{$^{\prime}$}}

\title{One or several populations of fast radio burst
sources?}

\author{M.~Caleb$^{1}$, L. ~G. Spitler$^2$ \& B. ~W. Stappers$^1$}

\begin{document}

\maketitle

\begin{affiliations}
 \item Jodrell Bank Centre for Astrophysics, the University of Manchester, Manchester, UK.
 \item Max-Planck-Institut f${\ddot{u}}$r Radioastronomie, Bonn, Germany.
\end{affiliations}

%%%%%%%%%%%%%%%
%% MAIN TEXT %%
%%%%%%%%%%%%%%%

\newpage

\begin{abstract}
To date, one repeating and many apparently non-repeating fast radio bursts have been detected. This dichotomy
has driven discussions about whether fast radio bursts stem from a single population of sources or two or more
different populations. Here we present the arguments for and against.
\end{abstract}

The field of fast radio bursts (FRBs) has increasingly gained momentum over the last decade. Overall, the FRBs discovered to date show a remarkable diversity of observed properties (see ref \cite{frbcat}, http://frbcat.org and Fig. 1). Intrinsic properties that tell us something about the source itself, such as polarization and burst profile shape, as well as extrinsic properties that tell us something about the source's environment, such as the magnitude of Faraday rotation and multi-path propagation effects, do not yet present a coherent picture. Perhaps the most striking difference is between FRB 121102, the sole repeating FRB\cite{Spitler}, and the more than 60 FRBs that have so far not been seen to repeat. The observed dichotomy suggests that we should consider the existence of multiple source populations, but it does not yet require it.

Most FRBs to date have been discovered with single-pixel telescopes with relatively large angular resolutions. As a result, the non-repeating FRBs have typically been localized to no better than a few to tens of arcminutes on the sky (Fig. 2). This relatively poor localization has made unambiguous associations with multi-wavelength counterparts and potential host galaxies challenging, thereby hindering a consensus for their origin. FRB 121102's repeating nature has permitted interferometric observations resulting in its localization to a dwarf galaxy at a redshift of z = 0.19 (ref. \cite{SriHarsh}) and deep multi-wavelength observations. Nevertheless, the exact nature of the source remains to be determined.

We can look to two fields in astronomy for historical guidance: radio pulsars and gamma-ray bursts (GRBs; see the Comment by S. R. Kulkarni in this issue) The radio emission from pulsars shows a dizzying range of properties. Much of this diversity correlates empirically with properties of the neutron star, such as its age, magnetic field and rotation period, but a detailed understanding of the origin of the radio emission cannot yet be fully explained on theoretical grounds. On the other hand, GRBs show us that the Universe is capable of producing transient events that are observationally similar but in fact are produced by at least two different source populations. Arguably the most important lesson to be learned from the study of GRBs is that identifying multi-wavelength counterparts and host galaxies was key to understanding their astrophysical nature.

The obvious case to make for multiple FRB populations is that the repeater has a different origin to the other known FRBs. Repeat bursts were discovered from FRB 121102 after only three hours of follow-up observations\cite{Spitler}, while the positions of other known FRBs have been observed for hundreds of hours or more with no repeats (see, for example, ref. \cite{Lorimer}). Unfortunately, when nothing is known about the astrophysical nature of these sources, the absence of repeat bursts is suggestive of, but cannot prove, the existence of multiple origins. Notably, FRB 121102 was discovered with the Arecibo Observatory, which is significantly more sensitive than any other radio telescope discovering FRBs to date. Potentially the source of FRB 121102 is intrinsically more active or closer than the other FRBs or its environment is more suitable for magnification of the bursts from plasma lensing\cite{Cordes}, and combined with Arecibo's raw sensitivity, this aided the rapid discovery of repetitions.

In any case, if all FRBs were repeaters and the statistics of the repeat rate were non-Poissonian, then initial bursts would be expected to be promptly accompanied by additional bursts, as in the case of FRB 121102 (ref. \cite{Niels}). In this scenario, prompt follow-up observations are more likely to yield repeating bursts. Since the detectability of pulses emulating a Poisson process is only affected by the total time spent observing the source, multiple short follow-up observations rather than long ones are more likely to yield repeat pulses. Also, FRBs detected by less sensitive telescopes would benefit from follow-up with higher sensitivity telescopes when looking for repeat pulses. Nevertheless, the absence of repeat pulses alone cannot rule out the existence of a single population. Detecting a typical wait time scale between pulses would provide a constraint on possible populations.

Are there observational indications for multiple populations beyond repetition? The observed durations of most FRBs are consistent with an intrinsically narrow burst widened by instrumental and propagation effects, but several FRBs do have multiple temporal components\cite{Champion}. The polarization properties of the eight FRBs with available polarization information show a mixed picture. For example, three of these bursts have linear polarization fractions greater than 80\% with little to no circular polarization, while one burst shows no linear polarization but a circular polarization fraction of 20\% (ref. \cite{frbcat}). But radio emission from Galactic neutron stars also shows a similar zoo of polarization properties. Whether the observed diversity in intrinsic properties is evidence of multiple populations is unclear and will require host galaxy identification and a much larger sample of FRBs to disentangle.

The environment in which an FRB resides could be key to understanding its origin. The propagation effects measurable in an FRB, which encode the physical properties of the media along the line of sight, could provide evidence for multiple populations. An FRB propagating through a magnetized, ionized medium will undergo Faraday rotation if the radio waves are linearly polarized, and the rate of rotation of the polarization vector (quantified by the rotation measure, RM) depends on the magnetic field strength and density of the medium. Much like the polarization properties of the known FRB population, the RMs also provide no coherent picture. Broadly speaking, the observed RMs fall into three groups: consistent with zero, a few hundred rad m$^{-2}$ (ref. \cite{frbcat}), or in the case of FRB 121102, roughly $10^{5}$ rad m$^{-2}$ (ref. \cite{Michilli}). If the Faraday-rotating material is causally connected to the FRB source, then reconciling a single population with a five orders of magnitude difference in the RM could be difficult. On the other hand, if the material causing the Faraday rotation is, by chance, along the line of sight, then a wide range of values would not be surprising. In any case the wide variation already observed suggests a range of different host environments.

As well as inspiring a large number of new observing campaigns and instrumentation, the discovery of FRBs has led to a large number of theoretical explanations that suggests that more than one class of FRB could indeed exist. Prior to the discovery of FRB 121102, the short duration and energetics of FRBs seemed to suggest that the triggering event was likely cataclysmic in some way, that is, it resulted in the destruction of the progenitor or companion. Examples include the merging of a pair of neutron stars or black holes (see, for example, ref. \cite{Totani}), the collision of a white dwarf or asteroid with a neutron star (see, for example, ref. \cite{Huang}), or the collapse of a heavy neutron star to a black hole (also known as a blitzar\cite{Falcke}). The challenge for these models is to create the diversity in burst shape, and polarization if it is intrinsic, with a single mechanism. While the blitzar model can potentially produce a couple of pulses, anything more complex doesn't fit with these models (see, for example, ref. \cite{Champion}). Perhaps these are the sources that should be targeted for repetition studies due to the hint of similarity to FRB 121102.

Models of a non-cataclysmic nature tend to concentrate on neutron stars because of their extreme gravitational and magnetic environments. It is possible that the FRBs represent some form of extreme pulsar (see, for example, ref. \cite{Popov}), or exceptional pulses from a giant-pulse-emitting pulsar (see, for example, ref. \cite{Wasserman}), although the former model does raise issues related to the total amount of energy available and the lifetime of the objects, which in turn has implications for the population of progenitors (that is, they may have to be multitudinous). The lack of an identified periodicity in the pulses from FRB 121102 also challenges these models. One of the first ideas for FRBs is that they are somehow associated with magnetars, which have energetics that are not limited by the spin-down energy of the neutron star\cite{Hurley}. An association with a high-energy source could help to determine whether this model is viable, and the more sporadic nature and emission across large fractions of the spin-phase in the radio may help to explain the lack of measured periodicity.

The inferred volumetric rate of FRBs is in the range 2,000--7,000 Gpc$^{-3}$ yr$^{-1}$ out to a redshift $z \approx 1$ (ref. \cite{Bhandari}). This is consistent with the volumetric rates of a range of transients like low-luminosity long GRBs, short GRBs, neutron-star neutron-star mergers and various types of supernovae such as core-collapse and type Ia\cite{Kulkarni}. In the simplest case, the repeater could belong to a different evolutionary phase of a given source population. For instance, rotating radio transients (RRATs) and FRBs could potentially be endpoints on a continuum across several orders of magnitude in luminosity. However in order to discern this enigmatic population, it is vital to localize the source to within a few arcseconds upon discovery, as repeating FRBs are rare, particularly if there is no afterglow or associated emission at any other wavelength that might help to reveal the location with sufficient precision.

The key question is: what observations should we carry out to settle the debate? Large samples of FRBs without precise localizations can provide an approximate solution to the nature of FRBs (such as those expected from CHIME and ASKAP in `fly's-eye mode') and potentially as cosmological probes. In particular, FRBs detected with polarization information and a temporal resolution not dominated by propagation effects would be particularly interesting. A sub-population of FRBs that is observed to repeat with relative ease next to a sub-population that is never observed to repeat, coupled with trends in burst profile, spectrum and polarization properties, could make a strong case for multiple populations.

But the crucial observation will be the properties of FRB host galaxies and multi-wavelength counterparts. FRB 121102 illustrated this clearly. The astrophysical origin scenarios were narrowed down through properties of the host galaxy and the persistent, compact radio source spatially coincident with the bursting source\cite{SriHarsh, Marcote}. Other than its existence, the only key observational clue to the origin scenario that the bursting source provided is its large rotation measure. So, while exploring the nature of FRBs themselves will provide important insights into their origin, it will be most effective when combined with information about host galaxy properties.

Interferometers with high sensitivities (such as MeerKAT and UTMOST) are expected to detect dozens of new FRBs with sufficient localization to be able to identify host galaxies. Arcsecond radio localizations can provide nearly all the insight required to determine the nature of the FRBs and their relations to other astrophysical transients. Additionally, the association of an FRB with a gravitational wave event due to a neutron-star merger would provide direct evidence of cataclysm in the absence of afterglows. However, it is evident from FRB 121102 that a multi-wavelength approach is necessary to understand FRB progenitors and the physics of their emission (see the Comment by Sarah Burke-Spolaor in this issue). The MeerLICHT optical telescope will co-point with the MeerKAT telescope, yielding simultaneous optical data for the radio observations and thus providing an opportunity for a detection of an optical counterpart to an FRB. If multi-wavelength counterparts exist, they can provide further insight into their progenitors and the full range of physical processes involved. However this is not strictly required to address the most simple questions about FRBs. Ultimately, localization along with the association of an FRB with a counterpart (for example, a superluminous supernova, active galactic nucleus or merger of neutron stars) is key to resolving the possible existence of multiple populations. 

\newcommand{\aap}{A\&A}
\newcommand{\aaps}{A\&AS}
\newcommand{\aj}{AJ}
\newcommand{\apj}{ApJ}
\newcommand{\apjl}{ApJL}
\newcommand{\apjs}{ApJS}
\newcommand{\araa}{ARA\&A}
\newcommand{\mnras}{MNRAS}  
\newcommand{\pasa}{PASA}
\newcommand{\pasj}{PASJ} 
\newcommand{\pasp}{PASP} 
\newcommand{\prd}{PRD} 
\newcommand{\nat}{Nature}
\newcommand{\sci}{Science}
\newcommand{\procspie}{Proceedings of SPIE}

\newpage

\bibliographystyle{naturemag}
\bibliography{sample}

\begin{addendum}
\item M.C. and B.W.S. acknowledge funding from the European Research Council (ERC) under the European Union's Horizon 2020 Research and Innovation programme (grant agreement no. 694745). L.G.S. acknowledges funding from the Max Planck Society.
 \item[Author Information] Reprints and permissions information is
  available at www.nature.com/reprints. The authors declare no
  competing financial interests. Readers are welcome to comment on
  the online version of the paper. Correspondence and requests for
  materials should be addressed to M.~C. (manisha.caleb@manchester.ac.uk) or  L. G. S (lspitler@mpifr-bonn.mpg.de).\\
\end{addendum}

\newpage

% \begin{figure}
% \includegraphics[scale=0.05]{./Fig1.png}
\begin{figure}[H]
  \includegraphics[scale=0.4]{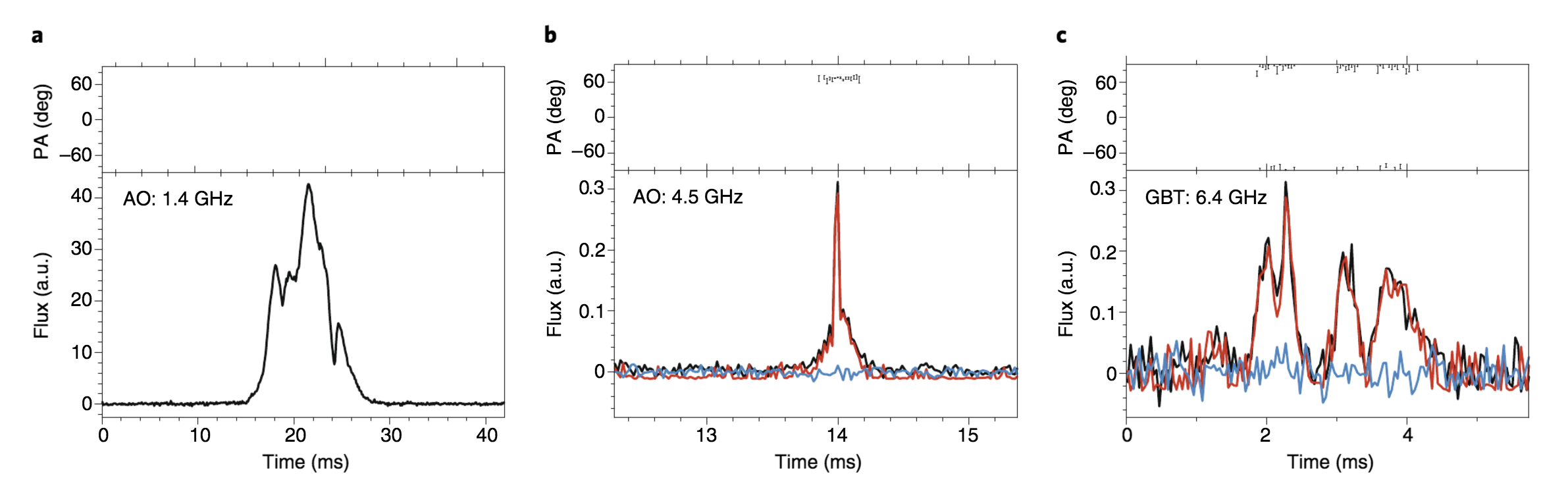}
\caption{The diversity of fast radio bursts. a–c, Three bursts from FRB 121102 are shown at three frequencies — 1.4 GHz (a), 4.5 GHz (b) and 6.4 GHz (c) — from the Arecibo Observatory (AO) (ref. \cite{Michilli} and J. W. T. Hessels et al., manuscript in preparation) and the Green Bank Telescope (GBT)\cite{Gajjar}. The 1.4 GHz burst from AO has at least seven distinct temporal components, and no measurable polarization, because the burst is completely depolarized at this frequency due to the source's large rotation measure. The other two higher-frequency bursts from FRB 121102 show that the emission is nearly 100$\%$ linearly polarized with a flat polarization position angle (PA) and that bursts exhibit a diversity of temporal behaviour at higher frequencies as well. The red and blue lines indicate linear and circular polarization profiles, respectively, while the black line is the total intensity. The absolute position angles plotted here are referenced to the centre of the observing band and therefore they differ from those in ref. \cite{Michilli} where they are referenced to infinite frequency.}
\end{figure}

\begin{figure}
\includegraphics[scale=0.4]{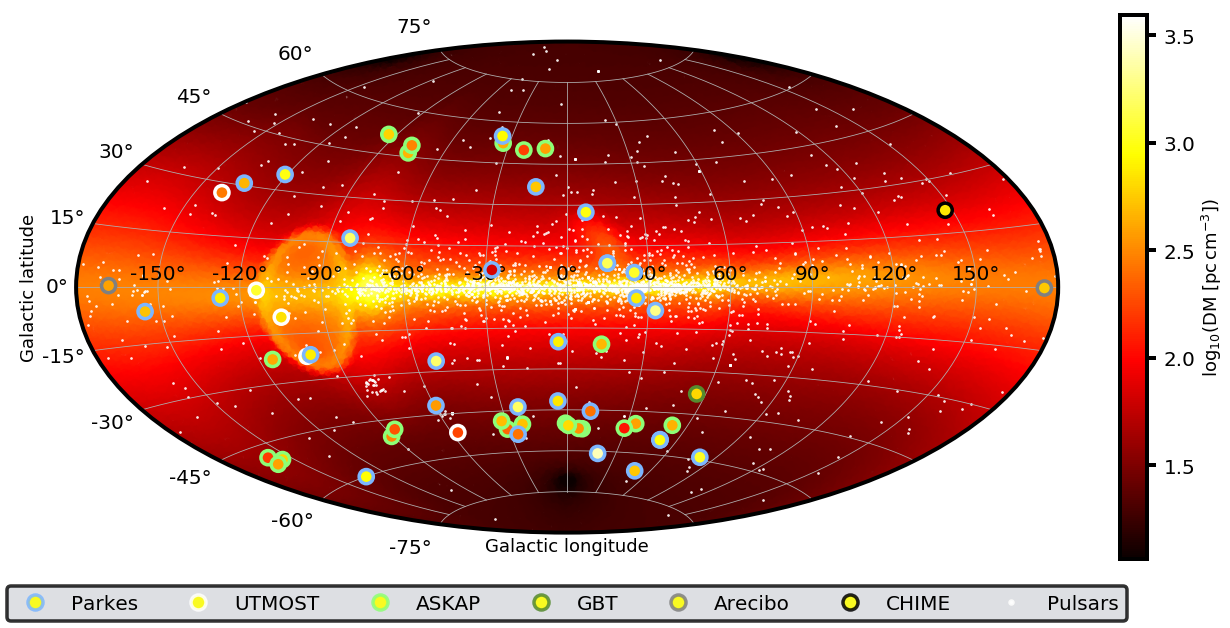}
\caption{The Galactic electron density map from the YMW16 model\cite{Yao} is shown in the background, giving the maximum Galactic dispersion measure along the given line of sight. Circles mark the positions of the 27 Parkes, 23 ASKAP, 5 UTMOST, 2 Arecibo, 1 GBT and 1 CHIME FRBs. The colour scale, which is proportional to the dispersion measure (DM), is common to both the background and the markers representing the FRBs discovered by the various telescopes. Any given FRB is seen to have a dispersion measure value much larger that the Galactic contribution at that sky position. Two of the Parkes FRBs have positions separated by 9$\arcmin$ and are not resolved in this figure. It should be noted that there are large biases in the spatial distribution due to large differences in sky coverage and survey depths. Credit: Laura Driessen, The Jodrell Bank Centre for Astrophysics, University of Manchester.}
\end{figure}

%% Here is the endmatter stuff: Supplementary Info, etc.
%% Use \item's to separate, default label is "Acknowledgements"

% \begin{addendum}
%  \item Put acknowledgements here.
%  \item[Competing Interests] The authors declare that they have no
% competing financial interests.
%  \item[Correspondence] Correspondence and requests for materials
% should be addressed to A.B.C.~(email: myaddress@nowhere.edu).
% \end{addendum}

\end{document}